\documentclass[pra,twocolumn,a4paper,showpacs,superscriptaddress]{revtex4}
\usepackage{amsmath}
\usepackage{amsfonts}
\usepackage{graphicx}
\begin{document}

\title{Distinctive subdynamic features of bipartite systems}  
\author{A. K. Rajagopal}
\affiliation{Center for Quantum Studies, George Mason 
University, Fairfax, VA 22030, USA.}
\affiliation{Inspire Institute Inc., McLean, VA 22101, USA.} 
\author{A. R. Usha Devi}
\affiliation{Department of Physics, Bangalore University, Bangalore-560 056, India.}
\affiliation{Inspire Institute Inc., McLean, VA 22101, USA.}
\author{R. W. Rendell}
\affiliation{Center for Quantum Studies, George Mason 
University, Fairfax, VA 22030, USA.}
\affiliation{Inspire Institute Inc., McLean, VA 22101, USA.}
\author{Michael Steiner}
\affiliation{Center for Quantum Studies, George Mason 
University, Fairfax, VA 22030, USA.}
\affiliation{Inspire Institute Inc., McLean, VA 22101, USA.} 

\date{\today}
\begin{abstract}
There are several important bipartite systems of great interest in condensed matter physics  and in quantum 
information science. In condensed matter systems, the subsystems are examined traditionally by using the Green 
function and mean-field-like methods based on the Heisenberg representation. In quantum information science, the 
subsystems are handled by composite density matrix, its marginals describing the subsystems, and the Kraus 
representation to elucidate the subsystem properties. In this work, a relationship is first established between 
the two techniques which appear to be distinct at first sight. This will be illustrated in detail by presenting 
the two methods in the case of the celebrated exactly soluable Jaynes - Cummings model (1963) of a two-state 
atom interacting with a one-mode quantized electromagnetic field. The dynamics of this system was treated in the 
Heisenberg representation by Ackerhalt and Rzazewski (1975). We present here the corresponding subdynamics using  
Kraus representation.  A relationship between the two approaches is established and the relative merits of the 
two techniques are discussed in elucidating the distinctive features of the subdynamics. The striking effects of 
interaction and entanglement are made transparent in two illustrative examples: the transformations that 
manifest in non-interacting number and spin representations due to interactions in their respective subspaces.  
\end{abstract}
\maketitle

\section{Introduction}
An important class of quantum many-body (QMB) systems involves at least two interacting species, for example, in 
condensed matter physics, electron - phonon, electron - photon systems are major bipartite systems and qubits 
shared by two parties A(lice) and B(ob) have played central roles in quantum information science (QIS). 
With the advent of recent advances in experimental techniques, an entirely different class of problems involving 
time-dependent phenomena, e.g., interacting bipartite system of atoms and radiation~\cite{4,5}, have opened up a 
new class of issues. The understanding of these systems lies in formulating the respective systems in seemingly 
different frameworks involving quantum theory: the Green function formalism based on the Heisenberg 
representation in QMB condensed matter for example~\cite{1} and the Kraus representation in QIS based on 
marginal density matrix formalism~\cite{2}. It is only recently the ideas of quantum entanglement and 
decoherence in QIS are making their inroads into understanding quantum phase transition in condensed 
matter~\cite{3}. 

The purpose of the present paper is to establish a relationship between the two frame works, 
that appear to be distinct at first sight, by exhibiting explicitly the same starting point and showing how they 
depart in detail.  Even though the total bipartite system evolves in 
time unitarily, the time evolution of the subsystems is quite different~\cite{2}. By introducing sub-dynamic 
Heisenberg operators, we find a novel dynamical description of the subsystem time evolution. The subsystem 
Heisenberg operators do not retain the pristine nature of the non-interacting particles, except at certain 
intervals of time. We illustrate this in the Jaynes-Cummings Model (JCM)~\cite{3a}  of interacting bipartite 
system of a two-level atom  and a single mode radiation field. 

Quantum-limited solid-state devices have shown that the model of an atom in a cavity, bathed in a laser field, 
serves as a generic theoretical model for describing these systems.  JCM 
 serves as an exactly solvable model of such two interacting 
disparate systems - a two-level system and an
electromagnetic field (EMF) (or more generally, a boson field) - under certain simplifying assumptions. 
These are the rotating wave approximation and weak coupling of the two-level atom with the field. 
JCM has recently been extended to strong coupling regime~\cite{6a}. 
It is worth pointing out the use of JCM in several other situations in the recent literature: 
entangled systems for controlling symmetric qubits in trapped ions~\cite{6b}, 
linear optics methods to generate symmetric qubits~\cite{6c}, collective atomic spin excitations~\cite{6d}, 
and stability of atomic clocks~\cite{6e}.

The Heisenberg scheme in JCM was given by Ackerhalt et al~\cite{7}. We  present here 
a relationship between this and the QIS approach of  non-unitary sub-dynamics.
While in the QMB framework, the effects of one species on the other are expressed in terms of various types of 
renormalizations, in QIS, these features are manifested differently via explicit subsystem operators that act in 
the space of the subsystem, details of which are elucidated by examining atom-photon coupled system in JCM.  
The dynamics of the subsystems of both the photon and the atom  exhibit consequent features of dynamical 
correlation.  Although the sub-dynamics of the photon and spin (2-state atom) are generally expected to be
complex for the interacting system, the Heisenberg sub-dynamic operators are shown here to
make explicit the physical processes, which deviate from the pristine dynamics.

This paper is organized as follows: in Sec.~II, the relationship of the Heisenberg and the Kraus 
representations of subsystems is established. In Sec.~III, detailed analysis on the subdynamics of JCM is 
given. In the final section IV, summary and concluding remarks are made.

\section{Relationship of Heisenberg and Kraus Representations of Subsystems}

Consider two systems A and B with the Hamiltonian (time-independent in this sequel) 
\begin{equation}
\label{1}	 
H=H_A+H_B+H_{AB}	
\end{equation}
as a generic interacting composite system. Here the first two terms represent the Hamiltonian of the two 
subsystems A and B and the third term is the interaction between them. The systems A and B are described by 
their corresponding Schrodinger operators, $a(0)$ and $b(0)$  respectively.
The unitary time evolution operator is given by 
\begin{equation}
\label{2}
U(t)=e^{-i\, t\, H}.
\end{equation}
The  Heisenberg  operators  $a_{H}(t)$ and $b_{H}(t)$ are then given by
\begin{eqnarray}
 \label{3}
 a_{H}(t)&=&U^\dag(t)\, a(0)\otimes I_B\, U(t) \nonumber \\ 
  b_{H}(t)&=&U^\dag(t)\, I_A\otimes b(0)\, U(t).
\end{eqnarray}
Here $I_A,\ I_B$   are respectively the unit operators in A and B subspaces. 
The time dependent Heisenberg operators (\ref{3}) are composite operators 
acting on the total Hilbert space of systems A,B.  
Due to unitary nature  of  time evolution, the pristine algebra of operators remain intact. 
Time derivatives of the Heisenberg operators  obey the equation $i\hbar\, \frac{{\rm d} O_H(t)}{{\rm 
d}t}=[H,O_H(t)]=H\, O_H(t)-O_H(t)\, H$.  Coupled equations containing complicated combinations of operators of 
both subsystems result, as the Hamiltonian contains terms representing the interaction between the two 
systems, thus leading to time dependent operators in the composite space. 

\subsection{ The Heisenberg Formulation}

The formal time evolution of the total system density matrix $\rho(t)$ in the Schrodinger representation is 
expressed in terms of the unitary evolution operator, (\ref{2}), in the standard way: 
\begin{equation}
\label{4}
\rho(t)=U(t)\, \rho(t=0)\, U^\dag(t).
\end{equation}
The Heisenberg representations (\ref{3}) of subsystem operators could be realized by considering the expectation 
values of $a(0),\ b(0)$: 
\begin{eqnarray}
\label{5}
{\rm Tr}_{A}{\rm Tr}_B\, [a(0)\otimes I_B\, \rho(t)]&=&
{\rm Tr}_{A}{\rm Tr}_B\,[a_H(t) \rho(0)]=\langle a_H(t)\rangle\nonumber \\ 
{\rm Tr}_A{\rm Tr}_B[I_A\otimes b(0)\rho(t)]&=&{\rm Tr}_{A}{\rm Tr}_B\,[b_H(t) \rho(0)]=\langle 
b_H(t)\rangle.\nonumber \\
\end{eqnarray} 
Here $\langle\ \rangle$ denotes average over the initial density matrix $\rho(t=0)$.  This approach to realize  
Heisenberg representation for the operators, opens up a natual way to investigate the corresponding 
time-dependent subsystem operators - defined exclusively on the subsystem spaces -  as will be shown in the next 
subsection. 

\subsection{The Kraus Formulation} 
In this formulation, one takes the initial density matrix $\rho(t=0)$ of the composite system to be  
uncorrelated i.e.,  given by a direct product of their initial density matrices: 
\begin{equation}
\label{6}
\rho(t=0)=\rho_A(0)\otimes\rho_B(0). 
\end{equation}
Substituting (\ref{6}) in (\ref{5}), we deduce time evolution of operators acting solely on the subspaces as 
follows: 
\begin{eqnarray} 
\label{7}
 \langle a_H(t)\rangle&=&{\rm Tr}_{A}{\rm Tr}_B\,[a(0)\otimes I_B\, \rho(t)]\nonumber \\ 
&=& {\rm Tr}_{A}{\rm Tr}_B\,[U(t)\, a(0)\otimes I_B\, U^\dag(t)\, \rho_A(0)\otimes \rho_B(0)]\nonumber \\ 
&=& {\rm Tr}_{A}[\tilde{a}(t)\, \rho_A(0)].
\end{eqnarray} 	
Here we have defined an operator that acts in the subspace of A only: 
\begin{eqnarray}
\label{8}
\tilde{a}(t)&=&{\rm Tr}_B\,[U^\dag(t)\, a(0)\otimes I_B\, U(t)\, I_A\otimes \rho_B(0)]\nonumber \\ 
&=&{\rm Tr}_{B}[a_H(t)\, I_A\otimes \rho_B(0)].
\end{eqnarray}
Similarly we define the operator,  $\tilde{b}(t)$ in the subspace B. 
We now establish the expressions for subsystem operators in terms of using Kraus formulation  for the  
corresponding non-unitary evolution of the subsystem density matrices. 
When the initial density matrix is of the un-correlated form,
$\rho(t=0)=\rho_A(0)\otimes\rho_B(0),$ the time-evolution of the marginal 
  density matrices assume the following structure:  
\begin{eqnarray}
\label{9}
\rho_A(t)&=&{\rm Tr}_B[U(t)\rho_A(0)\otimes \rho_B(0)]\nonumber \\ 
&=&\displaystyle\sum_i\, V_i(t)\rho_A(0)V_i^\dag(t), \\
\ & &  {\rm with }\ \ \displaystyle\sum_i\, V^\dag_i(t)V_i(t)=I_A,\nonumber   \\
 \label{10} 
\rho_B(t)&=&{\rm Tr}_A[U(t)\rho_A(0)\otimes \rho_B(0)]\nonumber \\ 
&=&\displaystyle\sum_j\, W_j(t)\rho_A(0)W_j^\dag(t) \\
\ & &  {\rm with }\ \  \displaystyle\sum_j\, W^\dag_j(t)W_j(t)=I_B. \nonumber
\end{eqnarray}
Using the Kraus representation (\ref{9}), (\ref{10}) for the time evolution of 
the subsystem density matrices  we get 
\begin{eqnarray*}
&{\rm Tr}_A{\rm Tr}_B\,[ a(0)\otimes I_B\, U(t) I_A\otimes \rho_B(0) U^\dag(t)]= 
{\rm Tr}_{A}[a(0)\, \rho_A(t)] \\ 
&\ \ \ \ \ \ \hskip 1in ={\rm Tr}_{A}[\tilde{a}(t)\, \rho_A(0)].
\end{eqnarray*}
 using which we obtain, 
\begin{equation}
\label{11}
\tilde{a}(t)=\displaystyle\sum_i\, V^\dag_i(t)\, a(0)\, V_i(t).
\end{equation}
Similarly a typical time dependent operator of the subsystem B has the form 
\begin{equation}
\label{12}
\tilde{b}(t)=\displaystyle\sum_j\, W^\dag_j(t)\, b(0)\, W_j(t).
\end{equation}
 Because of the non-unitary nature of the time evolution in the subspaces, the pristine algebra of the 
Schrodinger operators are not preserved. The dynamical aspects of the interactions are thus manifested in these 
new forms. It is important to note that in contrast to the Heisenberg representation which makes these 
Schrodinger operators operate in the total space independent of the initial state of the system, the subdynamic 
operators give more detailed information concerning the way in which the partner subsystem influences it. This 
helps us to understand how entanglement or correlation within the full many body system operates in contrast to 
the hitherto known renormalization ideas. Unlike in the composite Heisenberg representation,  time derivative of 
subsystem operators $\tilde{a}(t),\ \tilde{b}(t)$ in  (\ref{11}) or (\ref{12}) do not offer simple algebraic 
equations. 

In the next section we illustrate the above formalism by examining the well known JCM of interacting one-mode 
radiation field and a two-level atom. This is an exactly soluble model and thus offers a clear example where all 
the features derived formally in this section can be illustrated.

\section{Subdynamics in Jaynes-Cummings Model}

The Hamiltonian of the JCM is given by, 
\begin{equation}
\label{13}
H= \omega\,(a^\dag a+\frac{I_R}{2})\otimes I_A+\frac{\omega_0}{2}\, I_R\otimes \sigma_z 
 + g\, (a\otimes \sigma_+ +a^\dag\otimes \sigma_-).
\end{equation}
The first term in Eq.~(\ref{13}) is the Hamitonian of the one mode radiation with frequency $\omega$ ; the 
second term is that of the two-level atom with $\omega_0$  as the energy difference between its ground and 
excited states, and the last term is the dipole interaction between the radiation and the atom in the rotating 
wave approximation. The various operators here have the following properties: Radiation field is represented by 
the Schrodinger creation, $a^\dag$, and destruction, $a$, operators of quantized photons, obeying the 
commutation relations: 
  \begin{eqnarray}
  \label{14}
  [a,a^\dag]&=&I_R, \\ 
 I_R&=&\displaystyle\sum_{n=0}^\infty\, \vert n\rangle\langle n\vert,\ \ a^\dag a\vert n\rangle = n\, \vert 
n\rangle \nonumber 
  \end{eqnarray} 
The two level atom is described by the Pauli spin operators 
\begin{eqnarray}
\label{15}
\sigma_x&=&\left(\begin{array}{cc} 0 & 1 \\ 1 & 0 \end{array}\right),\  
\sigma_y=\left(\begin{array}{cc} 0 & -i \\ i & 0 \end{array}\right),\nonumber \\ 
\sigma_z&=&\left(\begin{array}{cc} 1 & 0 \\ 0 & -1 \end{array}\right),\ 
 I_A=\left(\begin{array}{cc} 1 & 0 \\ 0 & 1 \end{array}\right), \\ 
 \sigma_{\pm}&=&\frac{1}{2}\ (\sigma_{x}\pm i\ \sigma_{y}),\ \ \sigma_z\, \vert \uparrow\rangle 
= \vert \uparrow\rangle,\ \sigma_z\, \vert \downarrow\rangle 
= - \vert \downarrow\rangle. \nonumber       
\end{eqnarray}
Ackerhalt et. al.~\cite{7} express (\ref{13}) as $H=\omega N +C$, in terms of two commuting composite operators 
\begin{eqnarray}
\label{16a}
N&=& a^\dag a\, \otimes\, I_A + I_R\, \otimes \, \sigma_+\sigma_- \\ 
\label{16b}
C&=& -\frac{1}{2}(\omega-\omega _0)\, I_R\,\otimes\,\sigma_z +g\, 
(a\, \otimes \, \sigma_+ a^\dag\, \otimes \, \sigma_-).\nonumber \\
\end{eqnarray}

which are constants of motion. Heisenberg representation of 
photon, atom operators are then expressible in closed form~\cite{7}, 
as will be described later for purposes of comparing them with the 
subdynamic versions of the same operators. 

We first express the exact solutions of JCM in the following form~\cite{8}: 
\begin{eqnarray}
\label{17a}
H\, \vert 0,\downarrow\rangle &=&\Omega(0,\downarrow)\, \vert 0,\downarrow\rangle, \ 
  \Omega(0\downarrow)=\frac{1}{2}(\omega-\omega_0)=\frac{\bigtriangleup\omega}{2}, \nonumber \\
H\, \vert \phi(n,s)\rangle &=&\Omega(n,s)\, \vert \phi(n,s)\rangle, \nonumber \\ 
  \Omega(n,s)&=&\omega\, (n+\frac{1}{2})+(3-2s)\, \lambda_n,\\ 
 \lambda_n&=&\left[\left(\frac{\bigtriangleup\omega}{2}\right)^2+g^2\, (n+1)\right]^{1/2},  \\ 
 n&=&0,1,2,\ldots , s=1,2; \nonumber  \\ 
 \label{17b}
 \vert\phi(n,1)\rangle&=&\cos\theta_n\, \vert n+1, \downarrow\rangle +
 \sin\theta_n\, \vert n, \uparrow\rangle, \nonumber \\ 
 \vert \phi(n,2)\rangle&=&-\sin\theta_n\, \vert n+1, \downarrow\rangle +
 \cos\theta_n\, \vert n, \uparrow\rangle, \nonumber \\ 
\tan\theta_n&=&\frac{g\sqrt{(n+1)}}{\left(\frac{\bigtriangleup\omega}{2}\right)+\lambda_n}
\end{eqnarray}
The unitary time evolution operator $U(t)=e^{-it\,H}$ is given in terms of these solutions
\begin{eqnarray}
\label{18}
U(t)&=&\vert 0,\downarrow\rangle\, e^{-it\, \Omega(0,\downarrow)}\langle 0, \downarrow\vert \nonumber \\ 
& & \  +\sum_{n=0}^\infty\sum_{s=1,2}\, \vert\phi(n,s)\rangle  e^{-it\, \Omega(n,s)}\, 
\langle\phi(n,s)\rangle\vert
\end{eqnarray}
which can be expressed in terms of photon-atom corelation factors in a physically transparent manner as, 
  \begin{eqnarray}
\label{18a}
  U(t)&=&\displaystyle\sum_{n=0}^\infty\, \{\,e^{-i\omega\, t(n-\frac{1}{2})}\, v^*_{n-1}(t)\, 
  \vert n, \downarrow\rangle\langle n, \downarrow\vert  \nonumber \\ 
  && +e^{-i\omega\, t(n+\frac{1}{2})}
  \,[\, v_{n}(t)\,\vert n, \uparrow\rangle\langle n, \uparrow\vert \\
&&  -iw_{n}(t)\,\vert n+1, \downarrow\rangle\langle n, \uparrow\vert 
  -iw_{n}(t)\,\vert n, \uparrow\rangle\langle n+1, \downarrow\vert ]\nonumber \\ 
  \end{eqnarray}
Here the photon-atom correlation factors are given by, 
\begin{eqnarray}
v_n(t)&=&e^{-i\, \lambda_n\,t}\sin^2\theta_n+e^{i\,  \lambda_n\, t}\cos^2\theta_n,\nonumber \\ 
w_n(t)&=&\sin 2\theta_n\, \sin\lambda_n\, t.
\end{eqnarray}
  
\subsection { Heisenberg Operators in JCM} Ackerhalt et. at.~\cite{7} express 
the two basic equations of motion of the photon creation and the atomic 
flip up operators respectively in the forms, 
\begin{eqnarray}
\label{19}
\left(i\frac{\partial}{\partial t}+\omega\right)a_{H}^\dag(t)\hskip 0.15in&=&-g\, \sigma_{H+}(t) \\
\label{20}
\left(i\frac{\partial}{\partial t}+\omega+2\, C\right)\sigma_{H+}(t)&=&g\, a_{H}^\dag(t) 
\end{eqnarray}
Since the operator $C$ (see (\ref{16b})) is a constant of motion, exact Heisenberg operator solutions could be 
written down in terms of the linear combinations of the initial values of the operators. These exact solutions 
may be displayed in the following form to exhibit the composite nature of these solutions as follows:
\begin{eqnarray}
\label{21}
a_{H}^\dag(t)&=&  {\cal A}(t)\, a^\dag(t=0)+ {\cal B}(t)\, \sigma_+(t=0)\\
\label{22}
\sigma_{H+}(t)&=&  {\cal C}(t)\, \sigma_{+}(t=0)+ {\cal D}(t)\, a^\dag(t=0).
\end{eqnarray}
Here, the coeffecients $ {\cal A}(t),\  {\cal B}(t), \  {\cal C}(t),\  {\cal D}(t)$ are operator functions 
obtained by 
solving Eqs.~(\ref{19}), (\ref{20}): 
\begin{eqnarray}
\label{23}
{\cal A}(t)&=&e^{it\, \omega}\, \left(\frac{\hat r_+\, e^{it\,\hat r_-}-
 \hat r_-\, e^{it\,  \hat r_+}}{ \hat r_+- \hat r_-}\right)\nonumber \\ 
{\cal B}(t)&=& g\, e^{it\, \omega}\, \left( \frac{e^{it\,  \hat r_+}-
 e^{it\,  \hat r_-}}{ \hat r_+- \hat r_-}\right)\nonumber \\
{\cal C}(t)&=&e^{it\, \omega}\, \left(\frac{  \hat r_+\, e^{it\,  \hat r_+}-
 \hat r_-\, e^{it\,  \hat r_-}}{\hat r_+- \hat r_-}\right)\nonumber \\ 
{\cal D}(t)&=& -g\, e^{it\, \omega}\, \left( \frac{e^{it\, \hat r_+}-
 e^{it\,  \hat r_-}}{ \hat r_+- \hat r_-}\right)
\end{eqnarray}
where  $\hat r_{\pm}$ are given by 
\begin{equation}
\label{24}
\hat r_{\pm}=C\, \pm\, \sqrt{g^2(N-I_R\otimes I_A)
+\left(\frac{\omega-\omega_0}{2}\right)^2\,  I_R\otimes I_A}.
\end{equation}
Note that $\hat r_{\pm}$ are constant operators as they are specified by $C$, $N$ of (\ref{16a}), (\ref{16b}). 
It may be seen that at $t=0$, ${\cal A},\ {\cal C}$ are unit operators, while ${\cal B}, {\cal D}$ are zero. 
These have been used to re-examine the known physical properties of collapse and revivals by 
Narozhny et. al.~\cite{9}, by working out these operators in the representations employed in (\ref{17a}), 
(\ref{17b}). 

\subsection{Kraus formulation of subdynamic operators in JCM} 

In contrast to the Heisenberg operators of photon and atoms acting on the composite Hilbert space, we now give 
the Kraus formulation and display the operators in their respective subspaces. This is achieved by employing the 
expression for the time evolution operator, (\ref{18}) and then performing the procedures outlined earlier. 
Explicitly,  the reduced density matrix of the atom has the form: 

\begin{eqnarray}
\label{25}
\rho_A(t)&=&\displaystyle\sum_{N=0}^\infty\, W_{N\alpha}\rho^i_AW_{N\alpha}^\dag\nonumber \\
W_{N\alpha}&=&\langle N\vert U\vert \alpha\rangle, 
\end{eqnarray} 
where we have taken the initial state of the photon to be a pure coherent state 
$\rho_R(0)=\vert\alpha\rangle\langle \alpha\vert$, $\vert\alpha\rangle~=~\vert e^{-|\alpha|^2/2}\displaystyle
	\sum_{n=0}^\infty\frac{\alpha^n}{\sqrt{n!}}\,\vert n\rangle,$ $\alpha=|\alpha|\, e^{i\,\phi}$ 
with average number of photons given by $M=|\alpha|^2$	  and the photon number distribution is given by, 
$p(n)=\frac{e^{-M}M^n}{n!}.$ 

Similarly, the reduced density matrix of radiation is obtained as, 
  \begin{eqnarray}
  \label{26}
  \rho_R(t)&=&\displaystyle\sum_{s,s_1,s_2=\uparrow,\downarrow}\, (\rho^i_A)_{s_2s_1}\,V_{ss_2}\, 
  \rho^i_{R}\,V^\dag_{s_1s},\ \nonumber \\ 
  V_{ss_2}&=&\langle s\vert U\vert s_2\rangle. 
  \end{eqnarray}
From this we deduce the quasi- creation operator of the effective photon in the radiation subspace and the spin 
operators of the atom in its subspace:
\begin{widetext}
  \begin{eqnarray}
  \label{27}
  \tilde{a}^\dag(t)& =&e^{-i\,t\,\omega}\displaystyle\sum_{n=0}^\infty\left[ |n+1\rangle\langle n|\sqrt{n+1}\, 
A_n(t)   + |n+1\rangle\langle n-1|\, C_n(t) +|n\rangle\langle n|\, D_n(t)  \right]\\
  \label{28a}
 {\rm where} \ \  A_n(t)&=&\rho^i_{\uparrow\uparrow}\left(v_n(t)v^*_{n+1}(t)+w_n(t)w_{n+1}(t)\,
\sqrt{\frac{n+2}{n+1}}\right)
+\rho^i_{\downarrow\downarrow}\left(v_n(t)v_{n-1}^*(t)+w_n(t)w_{n-1}(t)\,\sqrt{\frac{n}{n+1}}\right) 
\nonumber  \\
 C_n(t)&=& i\,\rho^i_{\downarrow\uparrow}\left(w_{n}(t)v_{n-1}(t)\, 
\sqrt{n}-w_{n-1}(t)v_{n}(t)\sqrt{n+1}\right)  \\
\label{28c}
D_n(t)&=&i\,\rho^i_{\uparrow\downarrow}\left(  w_{n}(t)v^*_{n-1}(t)\,\sqrt{n+1}- 
w_{n-1}(t)v^*_{n}(t)\sqrt{n}\right). \nonumber   
\end{eqnarray}
\end{widetext}
At $t=0$  the quasi-photon operator $\tilde{a}^\dag(t=0)$ correctly reduces to the pristine photon creation 
operator 
\begin{equation}
\label{29}
  \tilde{a}^\dag(t=0)=\displaystyle\sum_{n=0}^\infty |n+1\rangle\langle n|\sqrt{n+1}.
\end{equation} 
The factor $A_n$  represents the suppression of single photon annihilation, while 
$C_n$, $D_n$ represent 2-photon and no-photon annihilations respectively. The precise nature of the photon-QPL 
will depend on the choice of the parameters 
and on the initial spin state. The 
standard commutation rule, $[a,a^\dag]=I_R$, does not hold for the photon-subsystem 
operators. These relations exhibit how the JCM interaction modifies the pristine field operators 
in important ways, thus leading to quasi-number representation in the radiation subsytem. 
In Eq.~(\ref{27}), the coefficients represent the effects of the photon - atom interaction in a clear fashion, 
reflecting the non-unitary feature of the Kraus transformation on the pristine photon operator, Eq.(\ref{29}). 
Similarly, the photon number operator is found to be
\begin{widetext} 
\begin{eqnarray}
\label{31}
\tilde{N}(t)&=&\displaystyle\sum_{s,s_1,s_2} (\rho^i_A)_{s_2s_1}\, V^\dag_{s_1s}\, (a^\dag a)\,V_{s,s_2}
\neq \tilde{a}^\dag(t)\tilde{a}(t) \nonumber \\ 
&=&\displaystyle\sum_{n=0}^\infty\,\left[\,\vert n\rangle\langle n\vert\, n 
+\vert n\rangle\langle n\vert\, (\rho^i_{\uparrow\uparrow}\, w^2_n(t)-
\rho^i_{\downarrow\downarrow}\, w^2_{n-1}(t)) -  i\,\rho^i_{\uparrow\downarrow}\,\vert n+1\rangle\langle 
n\vert\, w_n(t)v_n(t)
+  i\,\rho^i_{\downarrow\uparrow}\,\vert n\rangle\langle n+1\vert\, w_n(t)v^*_n(t)
\right]. \nonumber \\
\end{eqnarray} 
\end{widetext}
In Fig.~(1a) we display how far the operator structures of the photon, Eqs.~(\ref{27}), (\ref{31})
deviate from their pristine non-interacting versions in the subystem, for various values of the system 
parameters.  In the non-interacting case, $\langle\alpha\vert \tilde{a}(t)\vert\alpha\rangle=\alpha$,  
$\langle\alpha\vert \tilde{N}(t)\vert\alpha\rangle=\vert\alpha\vert^2$ 
and two sets of curves in Fig.~(1a) show the effects of interaction on these values as a function of $gt$.
\begin{figure}[h]
 \includegraphics*[width=3in,keepaspectratio]{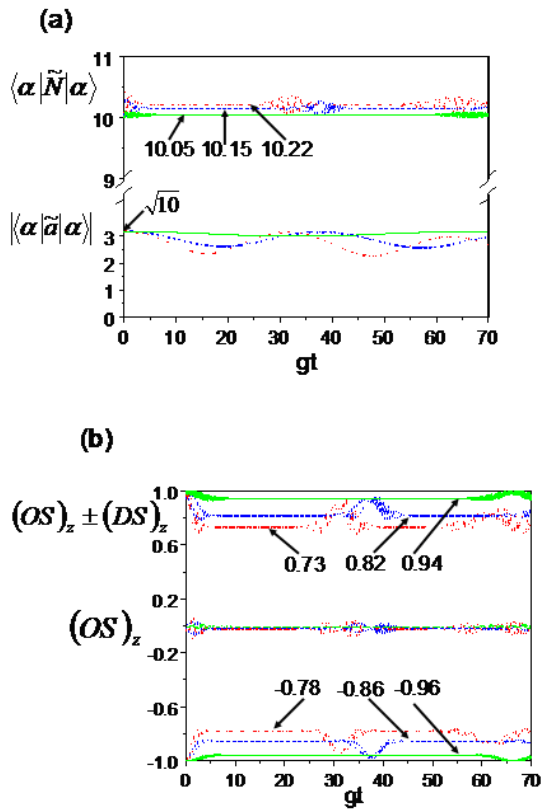}
\caption{(Color online)(a) Magnitude of $\langle\alpha\vert \tilde{a}(t)\vert\alpha\rangle$  and 
$\langle\alpha\vert \tilde{N}(t)\vert\alpha\rangle$
 vs. $gt$. The set of JCM parameters used are, Atom: $\rho_{\uparrow\uparrow}(0)=1,$
all others zero; Photon:   $N_{\rm mean}=\vert\alpha\vert^2=10;$
Interaction: dash-dotted line (red): $\Delta\omega/g=7.5$, 
dashed line (blue): $\Delta\omega/g=10$ and solid line (green): $\Delta\omega/g=20$.
Fig.~1\,(b) displays the offsets and eigenvalues of $\tilde{\sigma}_z(t)$ 
vs $gt$ for $N_{\rm mean}=10.$}  
\end{figure}
A similar calculation is now presented for the atomic operators in their Pauli spin form. Here the initial time 
representations of the pristine operators are given by
\begin{equation}
\label{32}
\sigma_+(t=0)=\vert \uparrow\rangle\langle \downarrow\vert=\sigma_-^\dag,\ 
\sigma_z(t=0)=\vert \uparrow\rangle\langle \uparrow\vert- \vert \downarrow\rangle\langle \downarrow\vert 
\end{equation}
and the corresponding subdynamic spin operators are given by
\begin{eqnarray}
\label{33}
 \tilde\sigma_+(t)&=& e^{i\omega\, t}\, \left(\vert\uparrow\rangle\langle\downarrow\vert\, S_1^+(t)
 + \vert\downarrow\rangle\langle\uparrow\vert \, S_2^+(t)\right. \nonumber \\
&& \ \ \ \ \ \ \left. + \vert\uparrow\rangle\langle\uparrow\vert\, S_3^+(t) 
 + \vert\downarrow\rangle\langle\downarrow\vert\, S_4^+(t)\right) \\
& =& [\tilde\sigma_-(t)]^\dag, \nonumber
 \end{eqnarray}
 with 
 \begin{eqnarray}
\label{34}
S_1^+(t)&=& \displaystyle\sum_{n=0}^\infty\,  p(n)\, v^*_n(t)v^*_{n-1}(t)\nonumber \\ 
 S_2^+(t) &=& \displaystyle\sum_{n=0}^\infty\, p(n)\, w_n(t)w_{n-1}(t)\, 
 \frac{\alpha^*\, \sqrt{n}}{\alpha\,   \sqrt{n+1}}\nonumber \\
S_3^+(t)&=& -i\, \displaystyle\sum_{n=0}^\infty\, p(n)\, v^*_n(t)w_{n-1}(t)\, \frac{\sqrt{n}}{\alpha} \\ 
S_4^+(t)&=&+i\, \displaystyle\sum_{n=0}^\infty\, p(n)\, v^*_{n-1}(t)w_n(t)\, 
\frac{\alpha^*}{\sqrt{n+1}}.
\end{eqnarray}
And,
\begin{eqnarray}
\label{35}
 \tilde\sigma_z(t)&=&  \vert\uparrow\rangle\langle\uparrow\vert\, S_1^z(t)
+\vert\downarrow\rangle\langle\downarrow\vert\, S_2^z(t) \nonumber \\
& +&
  \vert\uparrow\rangle\langle\downarrow\vert\, S_3^z(t)+
  \vert\downarrow\rangle\langle\uparrow\vert\, S_4^z(t),  
 \end{eqnarray}
 with
 \begin{eqnarray}
 \label{36}
S_1^z(t)&=& \left[1-2\, \displaystyle\sum_{n=0}^\infty\,p(n)\, 
w^2_n(t)\right]\\
S_2^z(t)&=&-\left[\,1-2\, \displaystyle\sum_{n=0}^\infty\,p(n+1)\, 
w^2_n(t)\,\right]\\ 
S_3^z(t)&=&-2i\,\displaystyle\sum_{n=0}^\infty\,p(n)\, w_n(t)v_n^*(t)\, 
\frac{\alpha}{\sqrt{n+1}}=S_4^{z*}(t)\nonumber \\
 \end{eqnarray}
Two highlighting features here are: (i) the pristine spin representations given in Eq.~(\ref{32}) get 
transformed in spin space due to photon - atom interaction, and (ii)  the eigenvalues of the new subdynamic 
effective spin operators thus have different eigenvalues and off-sets from their pristine values of +1, -1, and 
0 respectively. These are because of the non-unitary character of the Kraus transformation of the spin 
operators. the eigenvalues of  $\tilde{\sigma}_z(t)$ vs. $gt.$ are displayed in Fig.~(1b) for typical 
representative model parameters of JCM. These eigenvalues are given by $(OS)_z\pm (DS)_z$, with the off-set  
from zero is given by $(OS)_z=(S_1^z(t)+S_2^z(t))/2,$  and the dispersion is given by 
$(DS)_z=\left\{\left(\frac{(S_1^z(t)-S_2^z(t))}{2}\right)^2+\vert S_3^z\vert^2\right\}^{1/2}.$ 
In the non-interacting case, these are +1 and -1, with off-set equal to 0. 
The curves in Fig.~(1b) show these and, in addition, display the collapse and revivals. The collapses mimic 
constant values different from $\pm 1,$  depending on the parameters and the revivals occur at different times 
depending on the detuning. 
Similar results are obtained when other choices of parameters were made. 
The off-set being small and the dispersion different from +1,-1 
are the results of the non-unitary evolution governing the sub-dynamics. 

The above subdynamic features may also be interpreted in a complementary way:  
We have already shown how the pristine representations of spin and radiation 
get modified due to interaction and entanglement. We now 
exhibit this in terms of the constant of motion, $C$. 
Upon taking the expectation value of this operator over the total 
density matrix of the system, we obtain, 
$\langle n\rangle+\frac{1}{2}\, (\rho^i_{\uparrow\uparrow}-
\rho^i_{\downarrow\downarrow})=\langle\tilde{N}(t)\rangle+\frac{1}{2}\,\langle\tilde\sigma_z(t)\rangle$ , where
$\tilde{N}(t)$ and  $\tilde\sigma_z(t)$ are given explicitly by Eqs. (\ref{12}) and (\ref{14}) 
respectively. The back action on the photon number 
$\langle\tilde{N}(t)\rangle$  is such that it compensates the collapse
and revivals that occur in $\langle\tilde\sigma_z(t)\rangle$. 
Thus the photons and the atom  act 
in tandem in such a way as to maintain the constant of the motion.

\section{ Summary}

In this paper we have established a relationship between the traditional Heisenberg formalism for dealing with 
the dynamics of composite many-particle systems and the corresponding subdynamics based on Kraus representation. 
The latter brings out the effects of interaction and entanglement more directly than in the former description. 
The best known exactly soluble model of a bipartite system of a two-state atom interacting with a one mode 
photon field is treated in detail and the results are llustrated in Fig.~1. This figure displays how the photons 
and the two-state atomic states represented as effective spin half system are modified in novel ways beyond the 
already well-known results concerning them such as collapse and revivals. This is clearly brought out in Sec.~II 
where the JCM is described in detail in the two formalisms. We believe that such features are aspects of 
non-unitary sub-dynamics evolution of any multi-partite system when one examines how the respective pristine 
operators of the separate subsystem operators are modified in their respective subspaces. 

\end{document}